# Direct Visualization of Localized Vibrations at Complex Grain Boundaries

Eric R. Hoglund[1*||], De-Liang Bao[2||], Andrew O'Hara[2], Thomas W. Pfeifer[3], Md Shafkat Bin Hoque[3], Sara Makarem[1], James M. Howe[1], Sokrates T. Pantelides[2,4+], Patrick E. Hopkins[1,4,5°], and Jordan A. Hachtel[6♠]

[1] Dept. of Materials Science and Engineering, University of Virginia, Charlottesville, VA 22904, USA
[2] Dept. of Physics and Astronomy, Vanderbilt University, Nashville, TN 37235 USA
[3] Dept. of Mechanical and Aerospace Engineering, University of Virginia, Charlottesville, VA 22904, USA
[4] Dept. of Electrical and Computer Engineering, Vanderbilt University, Nashville TN 37235, USA
[5] Dept. of Physics, University of Virginia, Charlottesville, VA 22904, USA
[6] Center for Nanophase Materials Sciences, Oak Ridge National Laboratory, Oak Ridge, TN 37830, USA

||Contributed equally to the paper
*erh3cq@virginia.edu
+pantelides@vanderbilt.edu
°peh4v@virginia.edu
♠hachtelja@ornl.gov

## 1  First Paragraph

Grain boundaries (GBs) are a prolific microstructural feature that dominates the functionality of a wide class of materials. The change in functionality at a GB is a direct result of unique local atomic arrangements, different from those in the grain, that have driven extensive experimental and theoretical studies correlating atomic-scale GB structures to macroscopic electronic, infrared-optical, and thermal properties.[1–8] Here, we examine a $SrTiO_3$ GB using atomic-resolution aberration-corrected scanning transmission electron microscopy (STEM) and ultrahigh-energy-resolution monochromated electron energy-loss spectroscopy (EELS), in conjunction with density functional theory (DFT) calculations. This combination enables the direct correlation of the GB structure, nonstoichiometry, and chemical bonding with atomic vibrations within the GB dislocation-cores. We observe that nonstoichiometry and changes in coordination and bonding at the GB leads to a redistribution of vibrational states at the GB and its dislocation-cores relative to the bounding grains. The access to localized vibrations within GBs provided by ultrahigh spatial/spectral resolution EELS correlated with atomic coordination, bonding, and stoichiometry and validated by theory, provides a direct route to quantifying the impact of individual boundaries on macroscopic properties.

## 2 Introduction

The local atomic arrangement in a GB is different from that in the bounding grains to enable the accommodation of the misorientation of the corresponding lattices. Structurally, the local arrangements comprise dislocation-cores and structural units that repeat along the boundary. Chemically, dislocation-cores and other structural units are not always stoichiometric and may even feature complexions.[9] Together, the chemical and structural dissimilarities of GBs and grains lead to localized GB vibrations, which are of interest to many fields. For instance, in thermal transport[4–7,10] and infrared optics[4,8] phonon frequencies and lifetimes dictate they key aspects of the material response. Additionally, variations in the localized vibrations can significantly alter the free-energy landscape for functional materials[11–13] or increase entropic contributions to free-energy and influence phase transitions.[14–16]

For simple systems, theoretical predictions have been used to describe the unique relation between the structure and chemistry at a single GB to its local vibrational properties.[17–19] However, DFT calculations of GBs with small misorientation angle, especially in materials with complex structure and stoichiometry, become increasingly challenging, as they may necessitate excessively large supercells. On the other hand, for the analysis of complex GBs and other interfaces, aberration-corrected STEM has been a powerful tool that provides atomic-resolution high-angle annular-dark-field (HAADF) imaging,[3,16,20–24] while EELS core-loss and fine-structure analyses provides atomic-resolution chemical maps and bonding information.[25] While conventional STEM-EELS is limited to an energy-resolution of around 300 meV, modern electron monochromators are capable of 3 meV resolution.[26] More critically, atomic-scale spatial resolution can be achieved alongside this energy-resolution, enabling direct probing of vibrational properties[22,27–30], paving the way to correlate the vibrational properties of the GB directly to the chemical and structural properties.

Here, we examine a prototypical, highly complex, 10° [001] symmetric-tilt GB in $SrTiO_3$ using aberration-corrected and monochromated STEM imaging and EELS combined with DFT calculations, which are just manageable by computing power, for mutual validation. STEM imaging and core-loss EELS are used to directly measure the changes in stoichiometry and crystal structure that occur in and around the GB dislocation cores. The atomistic model of the GB is too large for conventional DFT relaxation, so the STEM/EELS data is used to facilitate the construction of the supercell, including local nonstoichiometries, that enables DFT calculations. Furthermore, energy-loss near-edge fine-structure (ELNES) analysis of the Ti and O edges at the GB together with DFT calculations reveal how the structural and chemical changes influence the bonding. Finally, monochromated EELS is used to probe the vibrations directly in the dislocation-cores (intracore vibrations) as well as between the dislocation-cores (intercore vibrations). Localized vibrational signatures are observed, which, through DFT calculations, are attributed to modes localized to GB dislocation-core atoms with unique environments.

## 3  Results

We begin by measuring the atomic structure and stoichiometry of the 10° GB using aberration-corrected STEM imaging and core-loss EELS spectroscopy. A HAADF image of this GB is shown in Figure 1(a) and Figure S1[31,32] and a zoom-in of the dotted white box is shown in Figure 1(b). Corresponding EELS-derived chemical maps are shown in Figure 1(d)-(g). Analysis of these data helped construct an atomistic model of the GB core (see details in Figure S8), optimized by DFT calculations, shown in Figure 1(c).

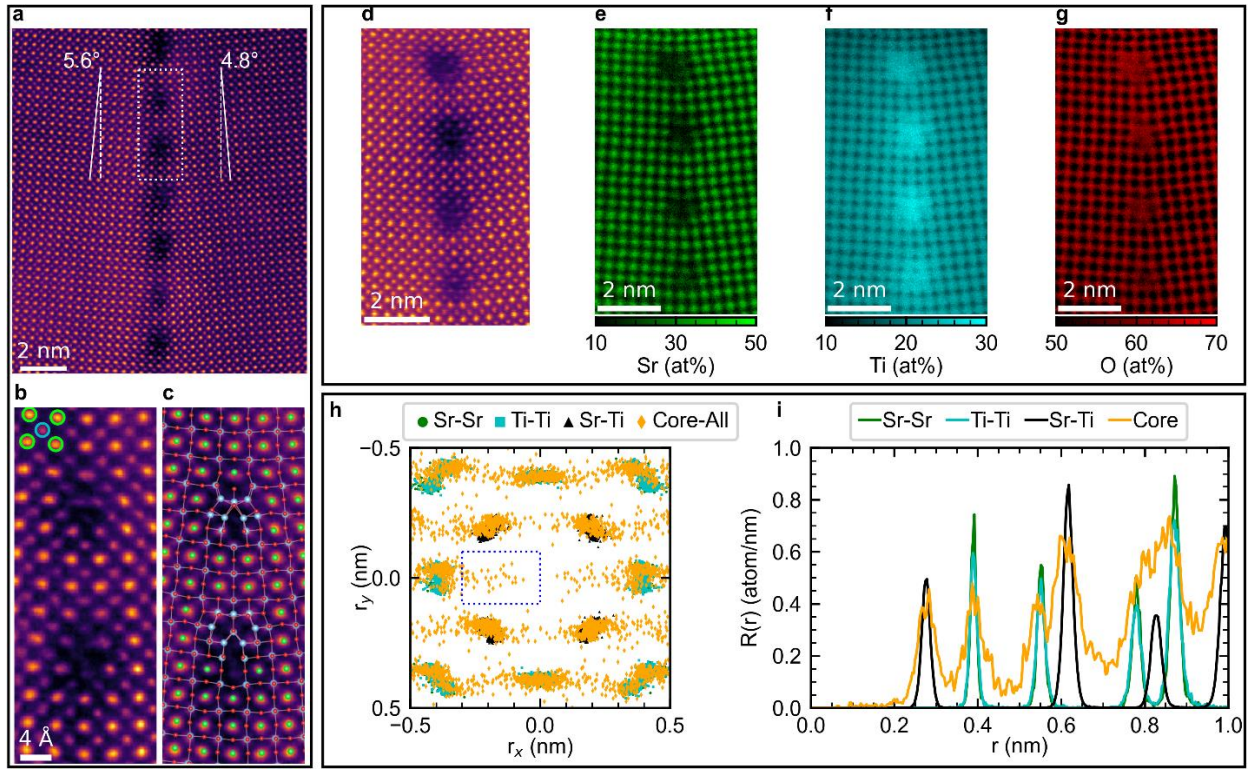

Figure 1. **Local atomic positions and stoichiometry at a 10° GB.** (a) HAADF image showing regularly spaced dislocations-cores at a GB. (b) An enlarged region of the white box in (a); (c) model of GB atomic positions based on the HAADF and EELS data. Annotated atom colors in (b) are green for Sr, cyan for Ti, and red for O. (d) Reference HAADF and (e-g) quantified compositions of (e) Sr, (f) Ti, and (g) O. (h) Two-dimensional nearest-neighbor distribution for (green) Sr-Sr, (cyan) Ti-Ti, and (black) Sr-Ti, shows well defined structure in the grains. Also shown is the (orange) nearest-neighbor distribution for atomic columns in the intracore regions created from all atomic columns in the image, which shows nearest-neighbors with smaller bond distances than the grain. The blue dotted box indicates atoms with shorter projected distance than in the grains, which indicates local changes in bond distance and/or coordination. (i) The radial distribution function formed from the nearest-neighbor distances in (h) further shows that the grains have well defined nearest-neighbor peaks while at the dislocations the probability is diffuse with non-zero minima and distances less than 0.2 nm.

The image in Figure 1(a) shows regularly spaced dislocations along the GB with markedly lower intensity than the grains, indicating a lower mass-density. Figure 1(a,b) demonstrates that the GB has different coordination than the grains. For instance, there are cations in the dislocation-cores with projected threefold coordination, whereas bulk $SrTiO_3$ cations have a projected fourfold coordination.

The EELS of Figure 1(d-h) show that the lower mass density at dislocation-cores corresponds to Sr-deficiency and Ti-enrichment, while the oxygen content at the cores is as in the grains, as also seen in Figure 1(c) and consistent with findings for other $SrTiO_3$ GBs.[31,32] We also note that the dislocation-core Ti and O signals are diffuse, compared with clear atomic columns in the grains, which suggests position variation of atoms in the projected direction.

To correlate the change in stoichiometry to the local coordination at the cores, we identify the positions of cation columns in the HAADF image, shown in Figure 1(a), and separate them into Sr and Ti sublattices in the grains, and atomic columns in the dislocation-cores (positions of atoms shown in Figure S2). With the defined sublattices, we can better understand local structure through the two-dimensional projected nearest-neighbor distribution and radial distribution function (RDF), shown in Figure 1(h,i). The grains show well-defined crystalline behavior within and between the Sr and Ti sublattices. In contrast, the nearest-neighbor distribution of atoms in the GB dislocation-cores show streaking along the direction normal to the boundary plane ($x$), without comparable streaking in the parallel plane ($y$), and have smaller projected nearest-neighbor distances than observed in the grains. The broader nearest-neighbor distribution is also reflected in the RDF in Figure 1(i) as broader nearest-neighbor peaks and a continuum background, illustrating that GB atoms have different coordination from grain atoms.

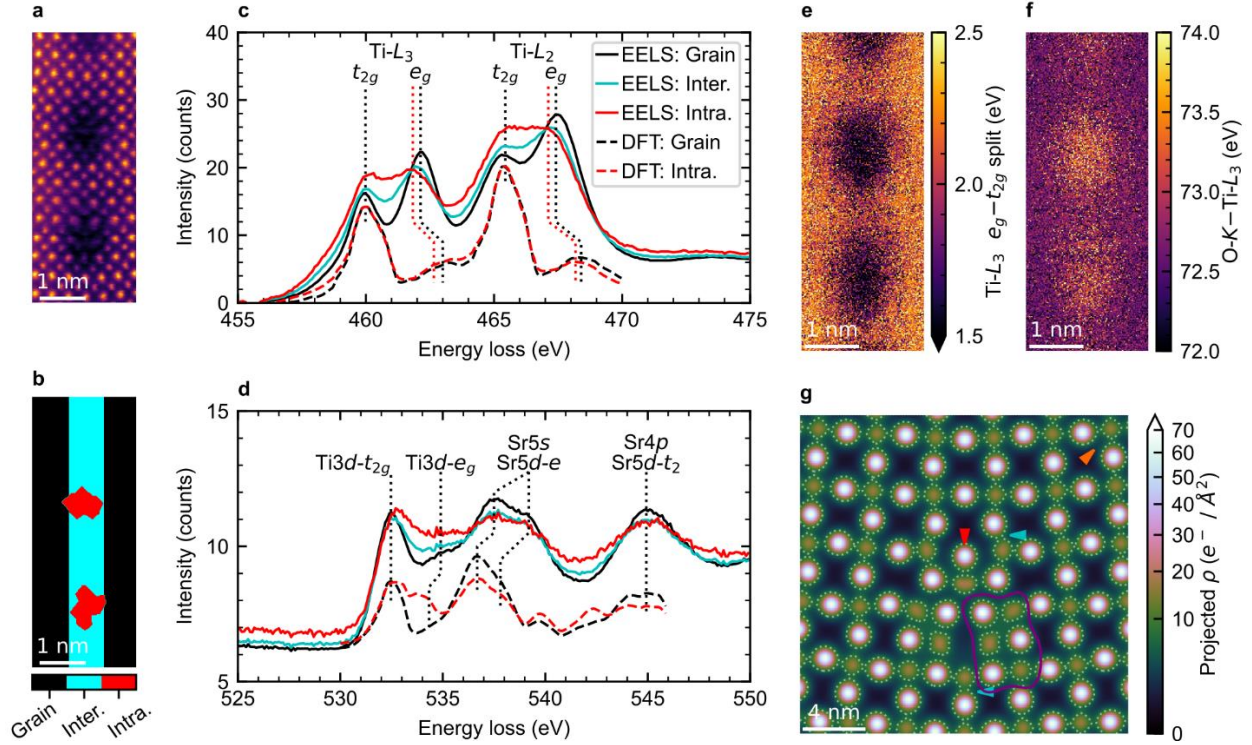

Figure 2. **Local bonding at a 10° GB.** (a) HAADF reference signal acquired during the STEM spectrum-image of the 10° GB. (b) Masks highlighting atom-columns located in the (red) intracore, (cyan) intercore, and (black) grain are used to create the average (c) Ti-$L_{23}$ edge and (d) O-$K$ edge core-loss spectra from the respective regions (the dashed curves are DFT-calculated spectra). The labels in (d) indicate the neighboring-cation orbital content of the final states; all of them also contain O $2p$. Z+1 DFT calculations from the grain and intracore region are shown in (c,d) as dashed lines and show similar Ti-$L_3$ $t_{2g}$-$e_g$ splitting and O-$K$ ELNES intensity changes as the experimental data. Maps of the (e) Ti-$L_{23}$ $t_{2g}$-$e_g$ splitting show the convergence from well-defined $t_{2g}$ and $e_g$ peaks into weakly split peaks. The difference between the O-K and Ti-$L_3$ onset energies (f) shows higher values at the dislocation-cores. (g) Projected valence-charge-density calculated using DFT. Regions of unique charge density are highlighted indicating (cyan arrow) increased orbital overlap, (red arrow) anisotropically bonded, and (purple outline) rock-salt packing are annotated and show orbital redistribution different than found in the (orange arrow) grains.

Another critical contribution to the characteristics of a GB is bonding within the dislocations. Since the compositional analysis revealed that the dislocation-cores comprise primarily Ti and O, we examine the spatial distribution of the Ti-$L_{23}$ and O-$K$ core-loss ELNES to understand the local bonding (Figure 2). Here, we consider three different regions: intracore, intercore, and the bulk SrTiO$_3$ grains on either side. Figure 2(a) shows a reference HAADF image simultaneously acquired during the ELNES acquisition, which we use to mask (red) intracore, (cyan) intercore, and (black) grain regions shown in Figure 2(b). The average Ti-$L_{23}$ and O-$K$ edge spectra from the regions shown in Figure 2(b) are shown in Figure 2(c,d), respectively. The Ti-$L_{23}$ spectra (Figure 2c) from the grains show well-defined $t_{2g}$ and $e_g$ peaks that result from the octahedral crystal-field splitting of Ti $d$-orbitals. The splitting of the $t_{2g}$ and $e_g$ peaks decreases to nearly one peak in the intracore spectrum for both the Ti-$L_3$ and Ti-$L_2$ edges. The intercore Ti-$L_{23}$ fine-structure is an intermediate of the grains and intracore, having smaller $t_{2g}$-$e_g$ splitting than the grains but more $t_{2g}$-$e_g$ splitting than the dislocation-cores, which shows that strain partially redistributes electronic states. A map of the Ti-$L_{23}$ $t_{2g}$-$e_g$ splitting in the GB, shown in Figure 2(e), reveals a gradual change from the grain to the GB, with the GB dislocation-cores having the largest change in electronic structure. In the O-$K$ edge, shown in Figure 2(d), the peaks related to Ti$3d$-O$2p$ orbital transitions in the grains, intercore, and intracore regions change intensity and approach a more continuous spectrum. The changes in splitting and redistribution are a direct result of the different coordination and varying geometric distortions observed in Figure 1. The onset-energies of the EELS are also different in each region suggesting different core-electron binding energies. The relative onset-energies between the Ti-$L_{23}$ and O-$K$ are mapped in Figure 2(f). Prior measurements of SrTiO$_3$ GBs showing similar stoichiometry and ELNES changes suggest a reduction from Ti$^{4+}$ to Ti$^{3+}$ at the dislocation-cores.[33–35] The intensity redistribution in both the Ti-$L_{23}$ and O-$K$ edges represent a redistribution in orbital energies, which is tied closely to the structural and stoichiometric changes shown in Figure 1.

To achieve deeper insights for the observed redistributions in electronic states, we performed DFT calculations on a DFT-relaxed atomistic GB model, a cropped region of which is shown in Figure 1(c) (see also Figure S8). First, we used the $Z$+1 approximation to calculate the ELNES features in the grains and at the intracore regions, shown in Figure 2(c,d). The calculated Ti-$L_{2,3}$ ELNES duplicates the decrease in the measured $t_{2g}$-$e_g$ splitting and the redistributions of states between and around the O-$K$ $t_{2g}$ and $e_g$ peaks. Likewise, agreement of intensity redistribution in the O-$K$ edge is found, demonstrating that the atomistic model and theoretical results are representative of the real GB dislocations.

We further examined the calculated GB valence-charge-density map, shown in Figure 2(g), to understand the redistribution of bonding orbitals at the GB. To focus on the TiO plane bonding, the map is calculated using the DFT real-space valence-charge density integrated from the TiO$_2$-plane to halfway between the SrO and TiO$_2$ planes (*i.e.* 1/2 to 1/4 of the unit cell, Figure S10 and Figure S11) along the z-direction. The typical bulk SrTiO$_3$ perovskite is highlighted in the top right corner with an orange arrow. The grains show well-separated isosurfaces bounding atoms, which is typical for of ionic materials. Three patterns of orbital redistribution relative to the grain are seen at the GB: regions of rock-salt packed TiO with charge density between ions [annotated in (g) with a purple box], regions of increased orbital overlap revealed by higher charge density between ion cores (cyan arrows), and regions where charge appears more localized on ions and anisotropically redistributed (red arrows). The broadening and decreased splitting of ELNES features at the GB relative to the grains reflects a complex redistribution of orbitals occurring at the GB in multiple ways. Therefore, the GB does not only exhibit different symmetries and concentrations, but the bonding between Ti and O atoms is stronger in some regions, weaker in others, and altogether irregular.

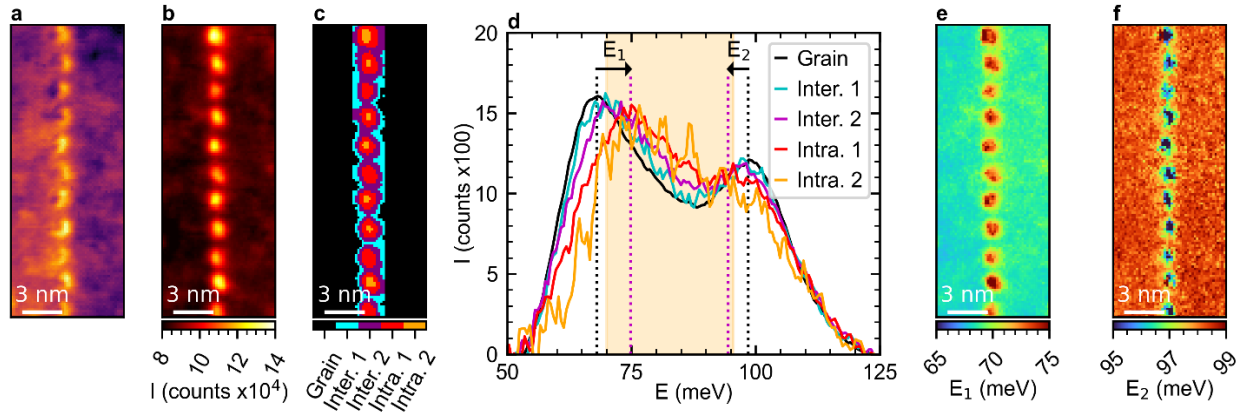

Figure 3. **Localized vibrational response of a 10° GB.** (a) aADF image and (b) total inelastic intensity from off-axis signal. (c) Sequence of masks generated from total inelastic intensity to selectively probe different degrees of intracore vs. intercore vibrational spectra. (d) Representative spectra of the masks in (c), showing a significant and continuous shift of the two peaks (labeled $E_1$ and $E_2$) as the region transitions from grain (black) to intercore (blue/purple) to intracore (red/orange). The gold spans in (d) marks the Reststrahlen band in the bulk material emphasizing a region where vibrational states redistribute to in the intracore signal. (e,f) Fit results for the peak energy of (e) $E_1$ and (f) $E_2$ showing localized blue-shifts of $E_1$ and similarly localized red-shifts of $E_2$.

Having shown that the structure, stoichiometry, and bonding at the GB and dislocation-cores are different from those in the grains, we now correlate these differences to localized changes in the atomic vibrations at the GB. To access highly localized vibrations at the 10° GB we use monochromated STEM-EELS in the off-axis geometry, as shown in Figure S6(b). The off-axis geometry results in an asymmetric annular dark-field (aADF) signal. The aADF image of the 10° GB, shown in Figure 3(a), exhibits contrast reversal relative to the HAADF images in Figure 1 and Figure 2. The higher aADF intensity at the intracore regions is likely from static-disorder. It is also observed in the total off-axis EELS signal shown in Figure 3(b), which further emphasizes the environmental disorder of atoms at the GB relative to the grains.[36]

To probe the grain, intercore, and intracore regions, we use the total EELS intensity to mask five distances from the dislocation-core, as shown in Figure 3(c). The average vibrational response from each characteristic region is shown in Figure 3(d). Significant differences between the grain, intercore, and intracore vibrational responses are readily observed. In the grain (black), there are two primary peaks at ~68 meV (labeled $E_1$) and ~99 meV (labeled $E_2$), which are consistent with known phonon modes from bulk $SrTiO_3$ originating from displacements of the Ti and O sublattices ($E_1$: second longitudinal optic (LO2)/third transverse optic (TO3) phonons, $E_2$: LO3 phonon).[37] However, as we transition from the grain (black) to the intercore region (blue/purple), the $E_1$ peak blue-shifts, while the $E_2$ peak red-shifts. The maximum shifts occur at the intracore region (red/orange), with fitted energies ~75 meV and ~94 meV, respectively. The peak fitting is applied to the entire spectrum-image, which is shown for $E_1$ in Figure 3(e) and $E_2$ in Figure 3(f) (fitting details shown in Figure S7). The high spatial-spectral resolution analysis of monochromated STEM-EELS enables us to see that the vibrational shifts match the spatial distribution of the aADF intensity in Figure 3 (a) as well as the different chemistry and fine-structure maps in Figure 1 and Figure 2, respectively. The agreement shows clear distinction between the grain, intercore, and intracore regions.

The peak shifts in the vibrational response can be attributed to the Ti-enrichment, coordination change, and bonding changes at the GB, as observed in Figure 1 and Figure 2. Similar localized effects are observed at antiphase boundaries where Ti-Ti neighbors lead to softening of the anti-ferroelectric optic modes, similar to observations at GBs in other materials.[37,38] These results demonstrate that a redistribution of vibrational states at a GB can be directly observed with monochromated EELS with the resolution necessary to correlate local structure, chemistry, and vibration characteristics of different regions at a GB.

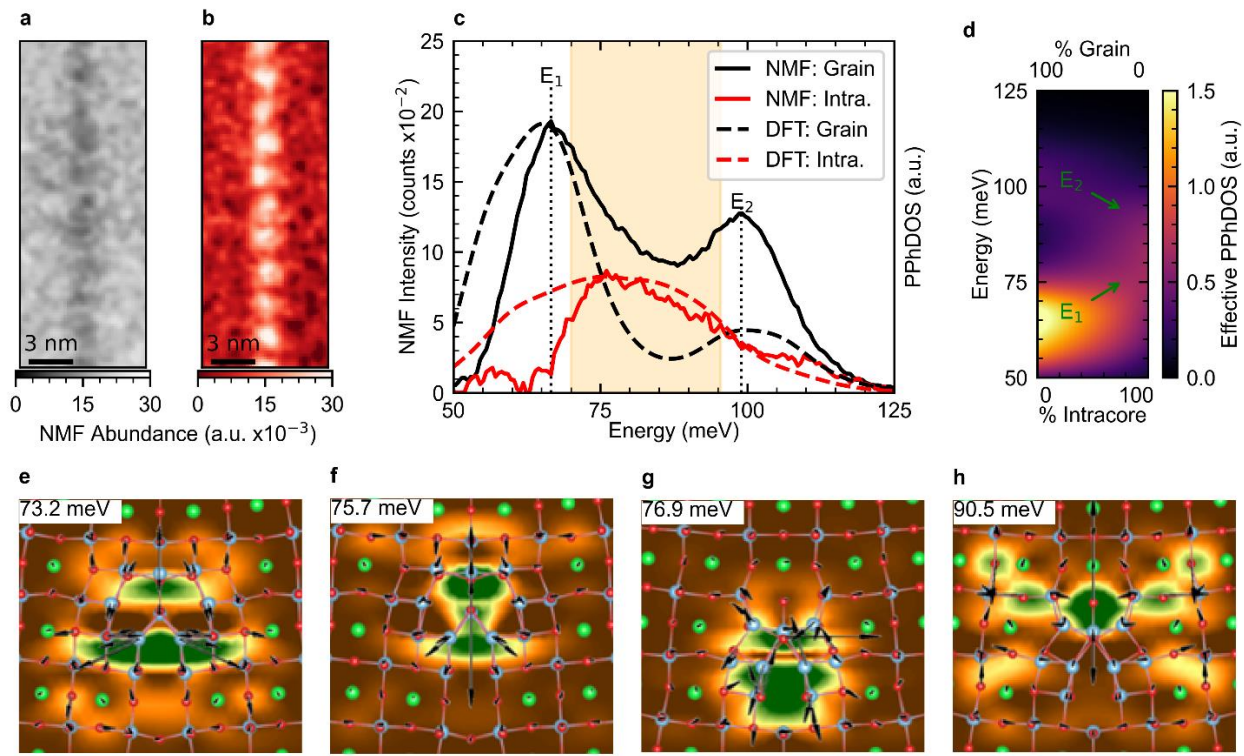

Figure 4. **Local GB vibrational modes in a 10° GB.** (a-c) Two component NMF analysis of off-axis GB spectrum-image. NMF maps show the decomposition of the spectrum-image data into a linear combination of (a) the grain signal and (b) the core signal. (c) The corresponding NMF spectra (solid curves) and DFT-calculated projected phonon density of states (PPhDOS) (dashed curves) exhibit overall good agreement. The grain spectra (black) match the raw EELS grain spectrum in Figure 3, while the core spectra (red) feature an asymmetric peak with a maximum at ~75 meV and a high energy tail. (d) Different linear combinations of the DFT GB-dislocation core and grain PPhDOS demonstrate that the measured vibrational spectra shown in Figure 3 can indeed be viewed as a combination of the intracore and bulk signals and varying concentrations (see also Figure S9). (e-h) Atomic displacements of vibrational modes localized within the GB cores at (e) 73.2, (f) 75.7, (g) 76.9, and (h) 90.5 meV.

To gain further insight into the vibrational shifts observed in Figure 3, we perform a non-negative matrix factorization (NMF) decomposition of the dataset. NMF is a well-established technique that can provide highly informative visualizations of hyperspectral EELS acquisitions, by reducing the dataset into a linear combination of spatial abundance maps of spectral features.[25] Figure 4(a-c) show a two-component NMF decomposition of the off-axis 10° GB EELS shown and analyzed in Figure 3, with the maps shown in (a) and (b), and the spectra shown in (c). We note that the grain component spectrum in Figure 4(c) is nearly identical to the grain vibrational response in Figure 3(d), and the core component spectrum in Figure 4(c) shows a peak at 75 meV, which is the energy that the $E_1$ peak shifts to in the intracore regions in Figure 3(d). Furthermore, the intensity of the grain component map shown in Figure 4(a) is not significantly reduced at the dislocation-cores, as in the average NMF abundance in the cores is only ~25% than the average NMF abundance in the grains (Figure S5). Conversely, the intensity of the core component map in Figure 4(b) is almost non-existant in the grains, with the average intensity of the core component reduced by ~85% compared to the intensity in the center of the intracore region. The critical implication of this finding, is that the measured signal (shown in Figure 3) is not the pure dislocation core signal, but rather a linear combination of the grain and core component spectrum at differing ratios, even with the increased localization of the off-axis EELS collection geometry. Therefore, the GB dislocations have vibrations associated with structure, nonstoichiometry, and bonding not present in the grain, but also have additional vibrations from local environments that share similarity with the perfect $SrTiO_3$ grain.

The nature of these localized core-component vibrational modes is unveiled using DFT calculations of projected phonon densities of states (PPhDOS). In Figure 3(c) we overlay the PhDOS of bulk $SrTiO_3$ as representative of the grains and the PPhDOS in the GB core with the EELS NMF components (see discussion in Figure S9 for how PPhDOS for each region is calculated). Both PPhDOS exhibit excellent agreement with the corresponding EELS NMF spectra, especially in the energy of the main peaks.

Moreover, the agreement helps us physicaly understand the peak shift occurring in Figure 3(d). Figure 4(d) captures the behavior by showing a transition from a pure grain signal to a pure intracore signal at different weighted intensities. The superposition of grain and intracore PPhDOS accurately reproduces the $E_1$ peak blue shift and $E_2$ peak redshift (an offset spectrum visualization of the effect is shown Figure S9).

We noted already that the GB-core NMF spectrum and the PPhDOS in Figure 4(c) feature a primary peak centered at ~75 meV that corresponds to vibrational modes within the gold-shaded Reststrahlen band of SrTiO$_3$, which ranges from 68 and 93.5 meV.[39] As a result, the pertinent modes are localized in the GB vicinity as they cannot propagate within the grains. We can also use the DFT calculations to visualize four such modes by plotting their displacement eigenvectors in Figure 4(e-g). The four modes, at eigenfrequencies 73.2 meV (e), 75.7 meV (f), 76.9 meV (g), and 90.5 meV (h), are representative of the intracore regions. The displacement amplitudes for all four modes are large at the center of the dislocation-core and small at the periphery, showing a high degree of localization in the intracore region. Furthermore, the localization of each mode is related to regions of different bonding shown in Figure 2(g). For example, the 73.2, 75.7, and 76.9 meV modes associated with Ti-O vibrations are localized to regions with coordination and bonding like a TiO rock-salt structure. The 90.5 meV mode is localized at the anisotropically bonded oxygen. We, therefore, identify the contribution of stoichiometry, highly irregular bonding, and local symmetry to the redistribution of vibrational modes at a GB.

## 4 Conclusions

The influence of GB structure, stoichiometry, and bonding on atomic vibrations is often left to simulations or inferred from a collection of boundaries in bulk samples. The combination of aberration-corrected STEM, monochromated EELS, and DFT calculations reveal that a GB and GB dislocation-cores have unique vibrations that are a direct result of local structure, stoichiometry, and irregular bonding. The ability to self-consistently measure and correlate coordination, chemistry, bonding, electronic states, and vibrational modes at the atomic scale enables the direct examination of single GBs that are complex in structure and composition. By unveiling the structure – chemistry – vibration relationships of GBs, their full impact on critical properties like infrared optical activity, thermal conductivity, heat capacity, and the vibrational entropy can be understood.

# Main Figure Captions

Figure 1. **Local atomic positions and stoichiometry at a 10° GB.** (a) HAADF image showing a string of dislocations-cores at a GB. (b) An enlarged region of the white box in (a); (c) model of GB atomic positions based on the HAADF and EELS data. Annotated atom colors in (b) are green for Sr, cyan for Ti, and red for O. (d) Reference HAADF and (e-g) quantified compositions of (e) Sr, (f) Ti, and (g) O. (h) Two-dimensional nearest-neighbor distribution for (green) Sr-Sr, (cyan) Ti-Ti, and (black) Sr-Ti, shows well defined structure in the grains. Also shown is the (orange) nearest-neighbor distribution for atomic columns in the intracore regions created from all atomic columns in the image, which shows nearest-neighbors with smaller bond distances than the grain. The blue dotted box indicates atoms with shorter projected distance than in the grains, which indicates local changes in bond distance and/or coordination. (i) The radial distribution function formed from the nearest-neighbor distances in (h) further shows that the grains have well defined nearest-neighbor peaks while at the dislocations the probability is diffuse with non-zero minima and distances less than 0.2 nm. ................................................................................................ 4

Figure 2. **Local bonding at a 10° GB.** (a) HAADF reference signal acquired during the STEM spectrum-image of the 10° GB. (b) Masks highlighting atom-columns located in the (red) intracore, (cyan) intercore, and (black) grain are used to create the average (c) Ti-$L_{23}$ edge and (d) O-$K$ edge core-loss spectra from the respective regions (the dashed curves are DFT-calculated spectra). The labels in (d) indicate the neighboring-cation orbital content of the final states; all of them also contain O 2$p$. Z+1 DFT calculations from the grain and intracore region are shown in (c,d) as dashed lines and show similar Ti-$L_3$ $t_{2g}$-$e_g$ splitting and O-$K$ ELNES intensity changes as the experimental data. Maps of the (e) Ti-$L_{23}$ $t_{2g}$-$e_g$ splitting show the convergence from well-defined $t_{2g}$ and $e_g$ peaks into weakly split peaks. The difference between the O-K and Ti-$L_3$ onset energies

## Acknowledgements

Monochromated EELS research was supported by the Center for Nanophase Materials Sciences, (CNMS), which is a DOE Office of Science User Facility using instrumentation within ORNL's Materials Characterization Core provided by UT-Battelle, LLC, under Contract No. DE-AC05-00OR22725 with the DOE and sponsored by the Laboratory Directed Research and Development Program of Oak Ridge National Laboratory, managed by UT-Battelle, LLC, for the U.S. Department of Energy. Aberration-corrected drift-corrected STEM imagining and core-loss EELS was supported by the Army Research Office, Grant No. W911NF-21-1-0119. Utilization of the Thermo Fisher Scientific Themis-Z STEM and Helios dual-beam focus ion beam instruments within UVa's Nanoscale Materials Characterization Facility (NMCF) was fundamental to this work. We thank Helge Heinrich for aiding in TEM sample preparation. Theory at Vanderbilt University (S.T.P., A.O. and D.-L.B.) was supported by the U.S. Department of Energy, Office of Science, Basic Energy Sciences, Materials Science and Engineering Directorate grant No. DE-FG02-09ER46554 and by the McMinn Endowment at Vanderbilt University. D.-L.B. was partially supported by the K. C. Wong Education Foundation of the Chinese Academy of Sciences. Calculations were performed at the National Energy Research Scientific Computing Center (NERSC), a U.S. Department of Energy Office of Science User Facility located at Lawrence Berkeley National Laboratory, operated under Contract No. DE-AC02-05CH11231.

# Author Contributions

E.R.H. and J.A.H. designed the experiments and performed all acquisition and analysis of STEM data. D.B., A.O., and S.T.P contributed all density-functional-theory calculations. T.W.P., S.B.H., S.M., and P.E.H. contributed acquisition, analysis, and understanding of the TDTR data. J.M.H. contributed his expertise and understanding of GB physics. E.R.H. wrote the manuscript and the supplemental information. All authors contributed to the direction and revision of the manuscript. Regarding reprints and permissions please contact E.R.H., J.A.H., P.E.H., or S.T.P. No financial or non-financial competing interests exist. Please contact any corresponding author for financial or non-financial questions.

# Data Availability

The datasets generated during and/or analyzed during the current study are available from the corresponding authors on reasonable request.

# 6 Methods

## 6.1 Samples and Sample Preparation

The 10° and 6° SrTiO$_3$ bicrystals were purchased from MTI Corporation. Plane view TEM samples were made using a Thermo Fisher Scientific Helios Dual Beam focused ion-beam. Initial milling and cleaning were performed at 30 kV, which was sequentially decreased until a finishing energy of 2 kV.

## 6.2 Electron Microscopy

Aberration-corrected STEM imaging and core-loss EELS were performed on a Thermo Fisher Scientific Themis-Z STEM operating at 200 kV with a 25 mrad convergence angle. Up to third-order aberrations were corrected for imaging and spectrum imaging. Drift-corrected imaging and analysis is explained further in Section 6.2.1 and core-loss EELS analysis is explained further in Section 6.2.2.

Vibrational EELS spectra were acquired at 100 keV using a Nion HERMES monochromated aberration-corrected dedicated STEM with a convergence angle of 30 mrad entrance aperture collection angle, and 0.525 meV/channel dispersions. Spectra were acquired as five-dimensional datasets, with one temporal ($t$), two spatial ($x$, $y$), one perpendicular momentum ($q$), and one energy-loss ($E$) dimension. Alignment, denoising, and fitting procedures are explained further in Section 6.2.3.

### 6.2.1 Aberration and Drift-Corrected STEM Imaging

STEM imaging was performed using a probe current of 20 pA. Two 1024x1024 px images were acquired at 90° relative scan rotations to compensate for linear and non-linear drift.[1] An example of the drift-correction for the image in Figure 1(a) is shown in Figure S1. Each image was acquired with a 64-μs dwell time. The image shown in Figure 1 was acquired with an 8-pm step size and the images in Figure S2(b-d) were acquired with a 16-pm step size.

Atomic columns within the grains were found and separated into Sr and Ti sublattices using atomap.[2] Atoms at the dislocation-cores were separated into a third sublattice. The border atoms of the dislocation-cores were defined by the first Sr atomic columns returning to the intensity of the grains' Sr atomic columns. The positions for atomic columns in each sublattice were refined by first finding the center of mass within a masked region 15% to the column's nearest neighbors. Further subpixel refinement was then iteratively performed by updating the nearest neighbor masked region and fitting two-dimensional Gaussians. The resulting atom positions are shown in

Figure S2. Nearest neighbor distributions were created using the cKDTree algorithm as implemented in SciPy with a radial cutoff of 3 nm.[3] Radial distribution functions were then calculated from the nearest neighbor distributions by binning the nearest neighbor distances.

### 6.2.2 Atomic Resolution Core-loss EELS

Core-loss spectrum images were acquired using a probe current of 100 pA and a spectrometer collection angle of 86 mrad. A Gatan K2 detector was used to acquire core-loss spectrum images with low signal-to-noise. The Ti-$L_{23}$, O-$K$, and Sr-$L_{23}$ edges were used for elemental quantification. Spectrum-images were then analyzed after singular-valued-decomposition denoising and in the raw state to provide low noise data for mapping while also representing the raw data, as shown in Figure 2 and Figure 3, respectively. Singular-valued-decomposition analysis is by nature a linear combination of singular spectra, so care needs to be taken when choosing the number of components in a system where peaks shift gradually over many pixels. In systems in which peaks shift dynamically, a compromise between noise and data features needs to be weighed carefully and then compared with the raw data and physics of the system. For further information on the denoising procedure, the reader is referred to Supplemental Section S2.1.

Mean spectra shown in Figure 2(d,e) are the averages of spectra from the grains, intracore, and intercore regions acquired with a 0.1 eV/channel dispersion. Here we define the intracore regions by a decreased intensity of the Sr-sublattice in HAADF images. The intercore regions between the intracore regions. The intracore regions were constructed by creating Voronoi cells of the atomic columns with SciPy[3]. Each Voronoi cells in the intracore regions were used to create a spatial mask, then the signal in each mask was averaged. The intercore region is then assigned as the remaining area ±1 unit cell from the boundary plane.

EELS from the full spectrum-images and Voronoi regions were fitted using least-square optimization as implemented in HyperSpy.[4] Fitting procedures for the stoichiometric mapping, ELNES mapping, and Voronoi mapping were performed in different manners and a full description of each can be found in Supplemental Section S2.2.

### 6.2.3 Vibrational EELS

The spectrum-image for the GB vibrations was obtained with a sequence of 30 individual spectrum images with dwell times of 10 ms per pixel with a 2D EEL spectrum (1028 x 130 pixels) captured for each SI pixel (total dataset size: 88 GB). The spectrum-image frame along the temporal direction is aligned using rigid registration to the total EELS signal, then averaged, providing good signal-to-noise with minimal drift-distortion. The zero of the energy axis is aligned using the full-width at half-maximum of the zero-loss or quasi-elastic peak. Delocalized dipole scattering decays as approximately $q^{-2}$ with high-$q$ signal localized to atomic length-scales.[5] Therefore, momentum transfer >35 mrad is integrated into a single spectrum for each pixel, providing highly localized signal, as shown in Figure S6(a,b). The 35 mrad cutoff is chosen to maximize signal-to-noise and the localization of the signal. The energy resolution measured as the full-width at half-maximum of the on-axis zero-loss peak is 13.64 meV and the resolution of the off-axis signal measured as the full-width at half-maximum of the quasi-elastic peak is 16.72 meV. The quasi-elastic peak at 0 meV has a large non-functional tail that was background subtracted by fitting an exponential to two windows at 45-55 and 130-140 meV, providing spatially comparable longitudinal-optic/third-transverse-optic and third-longitudinal-optic phonon peaks.

The region-of-interest comparisons are not denoised. For peak shift mapping, the spectrum-image is denoised using six components of a singular-valued-decomposition before background subtraction to minimize noise and increase the integrity of background fits. In the following, a fitting procedure is described, which is also shown in Figure S7 for a grain and intracore location. An intensity offset is measured between 275-280 meV, then a power-law function is fitted using two windows from 45-55 meV and 130-140 meV. The background is then fixed and two Gaussians to approximate the shifting of the two non-functional peaks seen in Figure 3(d). The first Gaussian is fitted from 62.5-72.5 meV with an upper bound set on the Gaussian center at 81 meV. The second Gaussian is then fitted from 94.5-105 meV with the Gaussian center bounded between 90-99 meV.

Before NMF a kernel filter is applied to the raw data and then background subtraction was performed. Kernel filtering enhances the signal-to-noise, providing more reliable background subtraction and less artifact in the NMF from the fitting procedure. NMF performed after background subtraction provides better sensitivity to changes in the optical phonon peaks and suppresses sensitivity to the high-intensity low-energy features.

### 6.2.4 Calculations

The DFT calculations were performed with the Vienna ab initio simulation package (VASP)[6] using the projected-augmented wave (PAW) method[7,8]. The local density approximation (LDA)[9] was adopted for the exchange-correlation functional based on the good performance of LDA on phonon calculations at the $\Gamma$ point[10]. The plane-wave basis energy cut-off was 500 eV for structural optimization and charge density calculations but 400 eV for the phonon calculations because of the calculational cost required by the big cell. The calculational supercell contains two cubic-STO domains with ~10° tilt angles separated by two parallel GBs. The GB consists of two alternating distorted cores (intracore) and deformed cubic STO lattices (intercore). To reduce the number of the modeling atoms, there are only two atomic layers along the *z* direction, i.e., one STO primitive cell in thickness, with a net number of 1130 atoms in the supercell, which is at the edge of practicality The lattice parameters of the supercell are 77.7, 42.7, and 3.87 Å along *x*, *y*, and *z* directions. The structure was relaxed until the interatomic forces are less than 0.02 eV/Å. The *k*-samplings are 1×1×3 for structural optimization and charge-density calculations, 1×1×1 for phonon calculations. The grain PhDOS was calculated using the cubic STO primitive cell, by projecting the eigenvectors in the (110) plane that is perpendicular to the electron beam. The intracore PhDOS were obtained using the supercell, by projecting 152-intracore atoms' vibrations in the (110) plane. The core-loss EELS was simulated using the Z+1 approximation, i.e., calculating the PDOS on one intracore oxygen atom that was replaced with fluorine and on one intracore titanium atom that was replaced with vanadium. In the theoretical spectra shown in Figure 2(c) and 2(d), "DFT-grain" means 100% grain, whereas "DFT-intra" means a sum of 30% intracore PDOS and 70% grain PDOS on corresponding atoms. The ratio was adjusted to optimize agreement with the experimental EELS.

## Methods References

## Supplemental Information Figure Captions

Figure S6. **Off-Axis EELS Collection Geometry and Resolution**. (a) Schematic for high spatial resolution vibrational EELS off-axis collection geometry. With a convergence semiangle of 30 mrad and a collection semiangle of 25 mrad, the beam is deflected by ~18 mrad such that the majority of the collection aperture accepts only high angle scattering events. This reduces the delocalized dipole scattering dominant in the bright field disk and emphasizes the localized impact scattering dominant at the higher angles to provide a high spatial resolution signal.[49,50] (b) Two-dimension display of the averaged hyperspectral dataset shown for vibrational EELS in the main text. Here, the x-axis represents energy loss (the dispersive axis of the spectrometer) and the y-axis represents the effective scattering angle (the non-dispersive axis of the spectrometer). The BF disk extends to ~30 mrad, and an additional 5 mrad is added to further suppress dipole scattering, and all higher angles are averaged to form the off-axis signal shown in the main text. (c) The effective energy resolution of the experiment can be determined by examining the full-width at

# S1 Supplemental Information: STEM Imaging

Images were drift-corrected using two orthogonal scan directions. An example of the drift correction for the image in Figure 1(a) is shown in Figure S1. The fast-Fourier-transforms in the insets of Figure S1(a,b) are skewed due to sample drift during serial imaging. The skewness of the image is removed after drift correction, as shown by the square reciprocal lattice in the fast-Fourier-transforms in the insets of Figure S1(c).

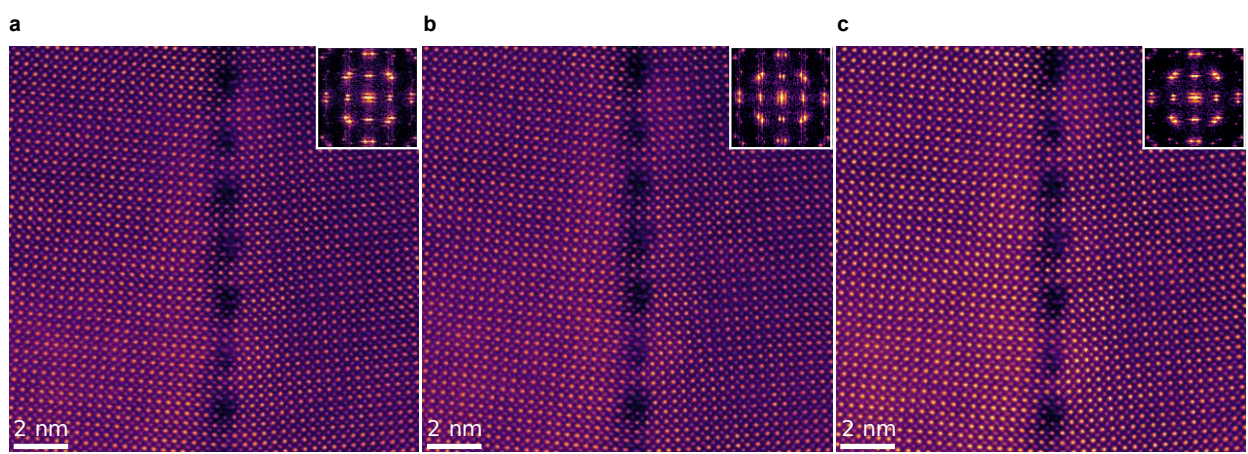

Figure S1. **Drift-correction of Aberration-corrected HAADF Image in the 10° GB.** Images acquired at (a) 0° and (b) 90° scan rotations with inset fast-Fourier-transforms showing a skewed reciprocal lattice. The image in (b) was rotated for better comparison with (a). (c) Image after drift correction.

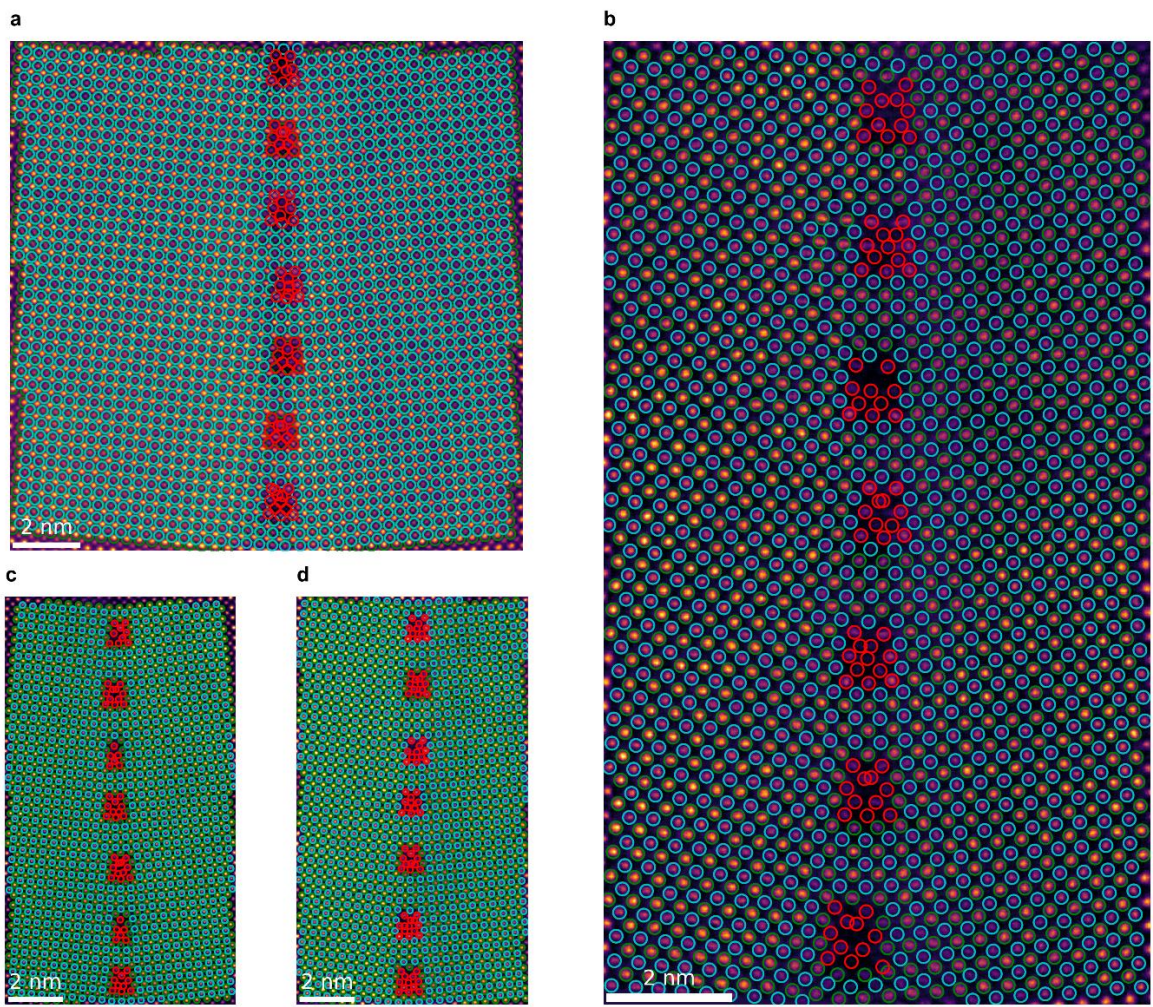

Figure S2. **Atom Positions in Four Drift-corrected HAADF Images.** (a-d) Drift-corrected HAADF images showing the (green) Sr, (cyan) Ti, and (red) core sublattices.

# S2 Supplemental Information: STEM Core-loss EELS

### S2.1 Denoising Spectrum Images

EELS core-loss excitations have a low scattering probability. Large atomic resolution spectrum images require short dwell times to minimize the impact of sample drift and electron beam damage. The combined low scattering probability and short dwell times for core-loss spectrum images therefore results in low signal-to-noise rations. Multivariate analysis, like singular value decomposition, provides a way to statistically denoise spectrum images by choosing a number of statistically relevant components and removing the rest.[1] Using discontinuities in the variance vs component number, otherwise known as a Scree plot, is a common method for identifying the number of components to include into a denoised model. However, if the spectral change(s) of interest is only present in a very small fraction of pixels or is a continuous shift over many pixels, then the spectral changes are not considered significant. In the case of GB dislocation-cores the changes in stoichiometry and structure are gradual, and the spectral shifts of the Ti-$L_{23}$ and O-$K$ are gradual. For example, a Scree plot for the Ti-$L_{23}$ and O-$K$ edge after background subtraction is shown in Figure S3. Two elbows are present in the explained variance ratio. The first elbow at component 6 includes most of the spectrum image features and will denoise the data. However, a continuous peak shift leads to a linear, slowly decaying explained variance ratio until another kink at component 110. Background, or improper background subtraction, can also lead to significant changes in the denoising process.

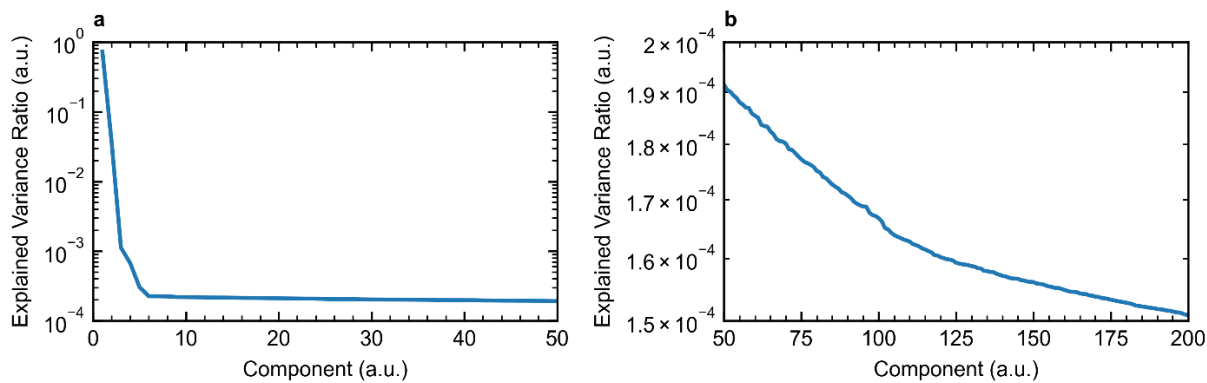

Figure S3. Scree plots for (a) 0-50 and (b) 50-250 components. (a) Shows the first elbow at component 4 and (b) shows the second elbow at component 110.

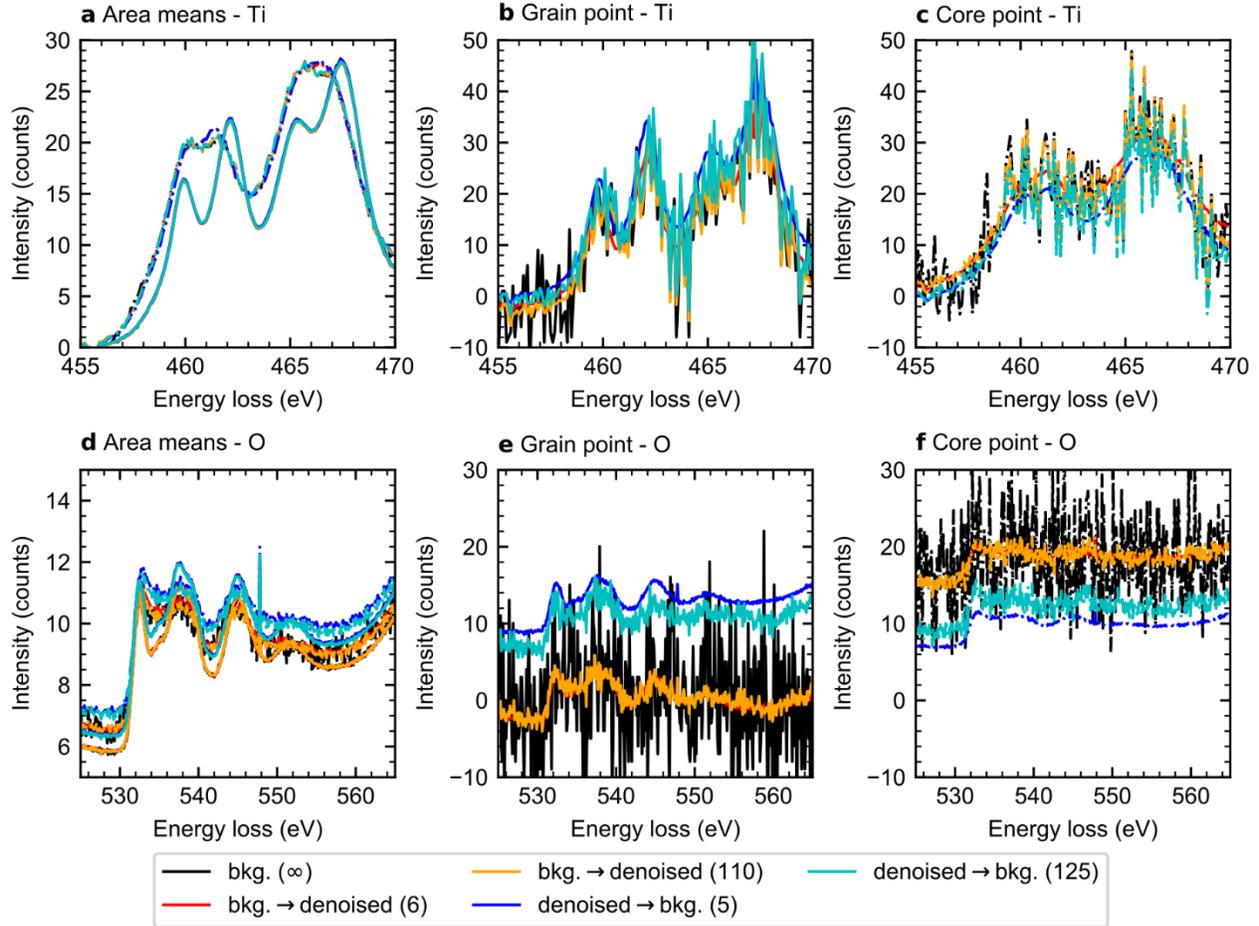

Figure S4. **SVD of TiO core-loss spectrum image of the 10° GB.** (a-c) Ti-$L_{23}$ edge (a) averaged from each region, and from individual points in the (b) grain and (c) GB. (d-f) O-$K$ edge (d) averaged from each region, and from individual points in the (e) grain and (f) GB. The colors of the lines correspond to the order of background subtraction and decomposition and the number of components, as indicated in the legend. In (a,c) the grain (solid) and dislocation-core (dot-dashed) signals overlap for each region because of the large number of averaged pixels.

To assure that our denoising process maintains the integrity of the original data, containing both bulk and defect spectral signatures, we plot denoised spectra constructed with multiple number of components in Figure S4, which is part of the denoising for the 10°-GB ELNES analysis in Figure 2. The combinations of background subtraction before and after denoising are also accessed. The area-averaged spectra from the grain and GB agree well with each other, because they contain many pixels and therefore represent a more significant portion of the pre-denoised data. On the other hand, the data from a single point in the grain or dislocation-core are less consistent as shown in Figure S4(b,c,e,f). Firstly, more intensity is distributed between the grain ELNES peaks when more components are included. Second, background subtraction before or after denoising influences significantly the total intensity of the O-$K$ edge. Fortunately, the size of the dislocation-cores is large enough that the difference between the grain and dislocation-cores is a significant portion of the spectrum image, so fewer components can be used and still be representative of most of the peak-shift trends while still reducing the noise. However, we note that some minor intensity redistribution is not included in the small component reconstructions here. Denoising can lead to significant oversight in other systems where peak shifts occur more abruptly over smaller fractions of a spectrum image.

### S2.2 Fitting ELNES Spectrum Images

EELS from the Voronoi regions were fitted with the following steps:

1. Fit Hartree-Slater core-loss edges between ELNES and EXELFS peaks (Ranges: 450 – 481, 482 – 526, 529 – 600).

2. Fit Ti-$L_{23}$ edges and three EXELFS Gaussians (Range: None – 450, 470 – 525). This step uses three Gaussians to approximate the EXELFS and ELNES + low-loss plural scattering.

3. Permanently freeze Ti-$L_{23}$ core-loss edges and EXELFS and do not fit for the remainder of the fitting procedure. Fixing the Ti-$L_{23}$ edge provides reproducible baseline for the core-loss edges based upon all post-ELNES intensity fit in step 2 and allows later for independent fitting of the ELNES intensity with minor accommodations made by the EXELFS peaks. Apart from reproducibility, removing the core-loss edges from the fit minimize fitting time and provide a more consistent procedure with the full spectrum image fit, which is explained later.

4. Fit the ELNES individually then together (Range: pre-edge – 470).

5. Freeze ELNES and unfreeze the EXELFS, then fit the EXELFS (Range: 470.5 – 525). Adjusts the EXELFS to accounts for plural added with ELNES.

6. Unfreeze the ELNES, then fit the ELNES and EXELFS together given that both sets should be near the correct values and plural scattering accounted for (Range: 450 – 525).

7. Freeze all Ti ELNES and EXELFS Gaussians

8. Add six Gaussians for the O-$K$ edge ELNES and one Gaussian for the Ti-$L_1$ edge ELNES. Then, independently fit the O-$K$ edge onset ELNES peak, Ti-$L_1$ edge ELNES peak, last O-$K$ edge ELNES peak, and second to last O-$K$ edge ELNES peak. Fit all O-$K$ and Ti-$L_1$ edge ELNES peaks together (Range: 525 – 570).

EELS TiO SI fitting:

1. Fit core-loss edges between ELNES and EXELFS peaks

2. Fit Ti edge and EXELFS three Gaussians (Range: None – 450, 470 – 525). Use the Voronoi grain EXELFS Gaussian fits as an initial condition.

3. Freeze core-loss edges and EXELFS (No more edge fits)

4. Add four ELNES Gaussians using the Voronoi grain EXELFS Gaussian fits as an initial condition. Then, fit only the ELNES Gaussians.

5. Freeze the core-loss edges and fit all ELNES and EXELFS peaks

## S3 Supplemental Information: STEM Vibrational EELS

While conventional STEM-EELS is limited to an energy-resolution of around 300 meV due to the cold field emission source, modern electron monochromators can improve the energy resolution to as high as 3 meV.[2] More critically, atomic-scale spatial resolution can be achieved alongside this energy-resolution when an off-axis geometry is used to suppress dipole scattering and enhance impact scattering.[3–5,5–7] The combined energy and spatial resolution pave the way to correlate the vibrational properties of interfaces, grain boundary, dislocations, and other low-dimensional heterogeneities directly to the chemical and structural properties. For off-axis EELS, the beam is deflected such that the principal optic axis is no longer collected by the EELS entrance aperture. As a result, the signal collected from the annular detector no longer collects radially symmetric annular signal, resulting in the asymmetric annular darkfield image (aADF) shown in the main text.

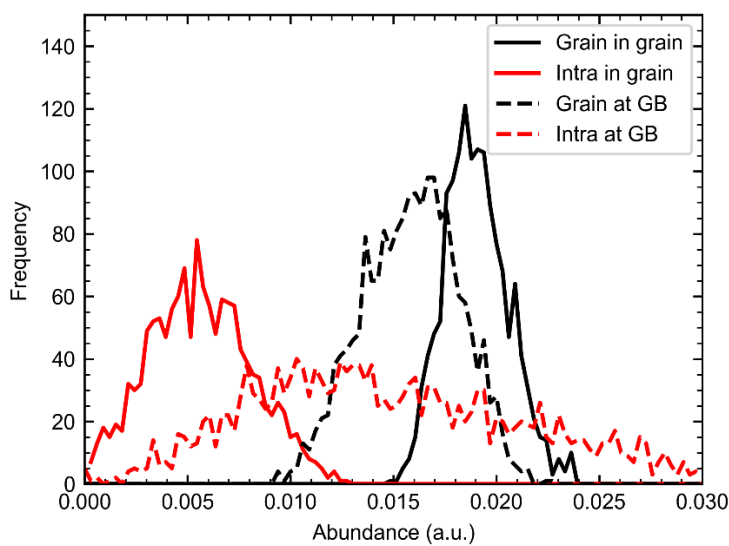

Figure S5. Histograms of NMF abundance maps in Figure 4(a,b). Distributions in the grain (solid) are from x-positions >1.5 nm from the GB center position. Distributions at the GB (dashed) are from -1.5 to 1.5 nm from the GB center position.

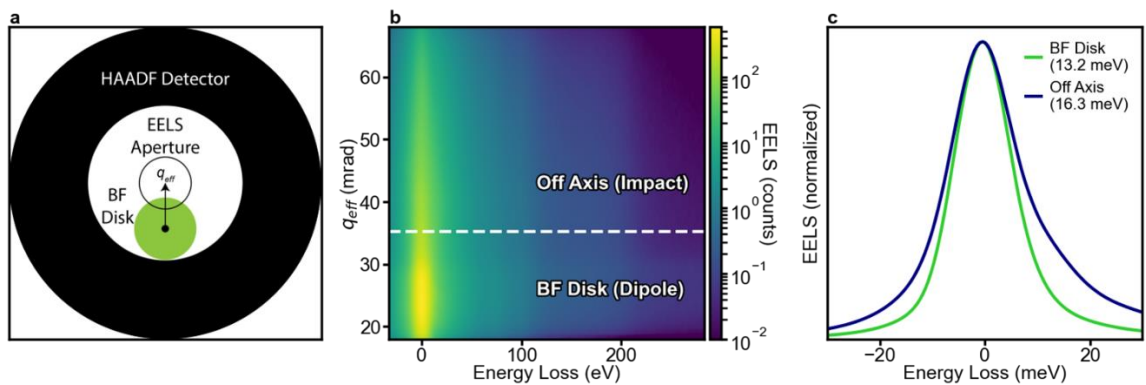

Figure S6. **Off-Axis EELS Collection Geometry and Resolution**. (a) Schematic for high spatial resolution vibrational EELS off-axis collection geometry. With a convergence semiangle of 30 mrad and a collection semiangle of 25 mrad, the beam is deflected by ~18 mrad such that the majority of the collection aperture accepts only high angle scattering events. This reduces the delocalized dipole scattering dominant in the bright field disk and emphasizes the localized impact scattering dominant at the higher angles to provide a high spatial resolution signal.[7,8] (b) Two-dimension display of the averaged hyperspectral dataset shown for vibrational EELS in the main text. Here, the x-axis represents energy loss (the dispersive axis of the spectrometer) and the y-axis represents the effective scattering angle (the non-dispersive axis of the spectrometer). The BF disk extends to ~30 mrad, and an additional 5 mrad is added to further suppress dipole scattering, and all higher angles are averaged to form the off-axis signal shown in the main text. (c) The effective energy resolution of the experiment can be determined by examining the full-width at half-maximum (FWHM) of the elastic scattering peak. For the BF disk the FWHM is measured to be 13.2 meV, while in the off-axis which is more sensitive to lower energy phonons the energy resolution is reduced to 16.3 meV.

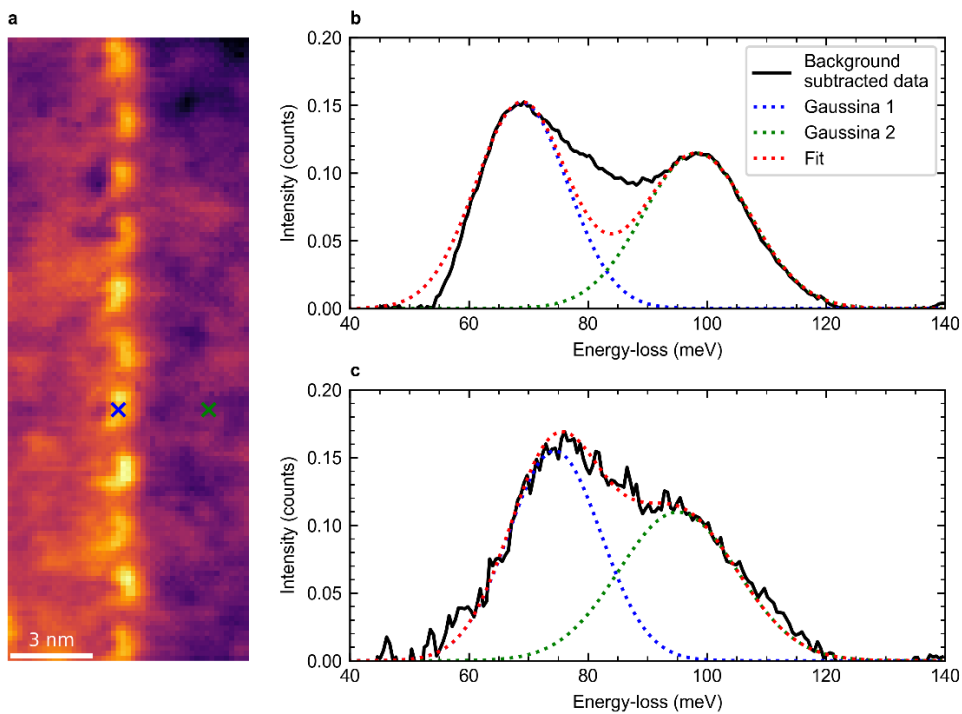

Figure S7. **Phonon peak shift fitting example.** (a) aADF image indicating a (green) grain and (blue) GB position chosen to show representative fits. Panel (b) show the grain fits and panel (c) shows the GB fits.

## S4 Supplemental Information: DFT

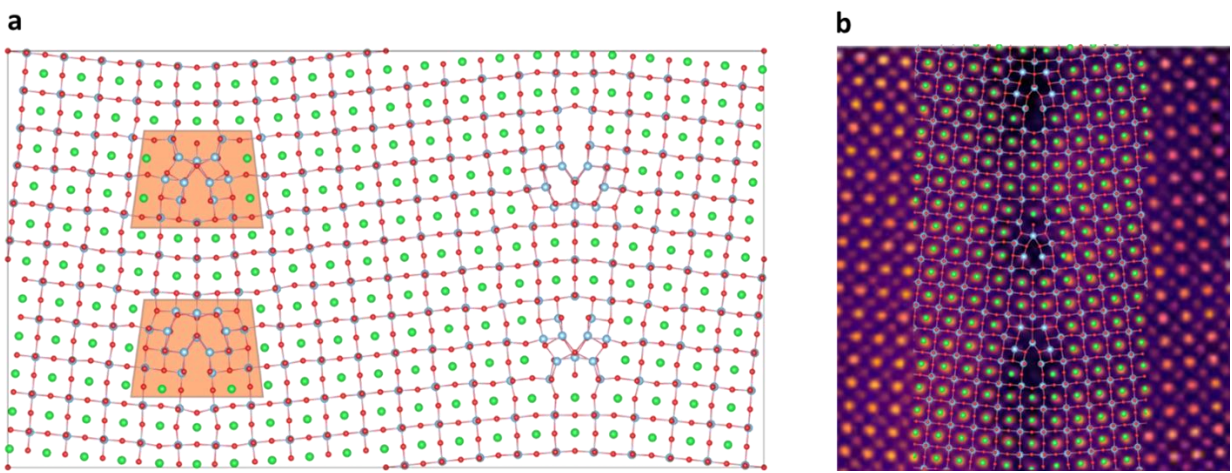

Figure S8. Construction of DFT Supercell. A model of the 10° low-angle GB is shown in (a). The dislocation-cores were constructed considering the brightness, i.e., masses of atoms, contrast in the atomic-resolution HAADF image (Figure 1(b)) and the chemistry distribution (Figure 1(e-g)), i.e., depletion of Sr and enrichment of Ti at the intracore. The orange masks in (a) highlight the alternating dislocation-cores. In the dislocation-cores region, the depleted Sr atoms were re-placed by Ti atoms. Moreover, additional Ti atoms were distributed in the middle of SrO and TiO atomic planes, together with surrounding O atoms, forming rock-salt-TiO-like configurations. The super-cell must contain two anti-parallel GBs to satisfy periodic boundary conditions. The distance between two anti-parallel GBs is ~3.9 nm. (b) Composition of large-scale atomic-resolution HAADF and DFT-optimized GB model, showing satisfactory agreement with each other.

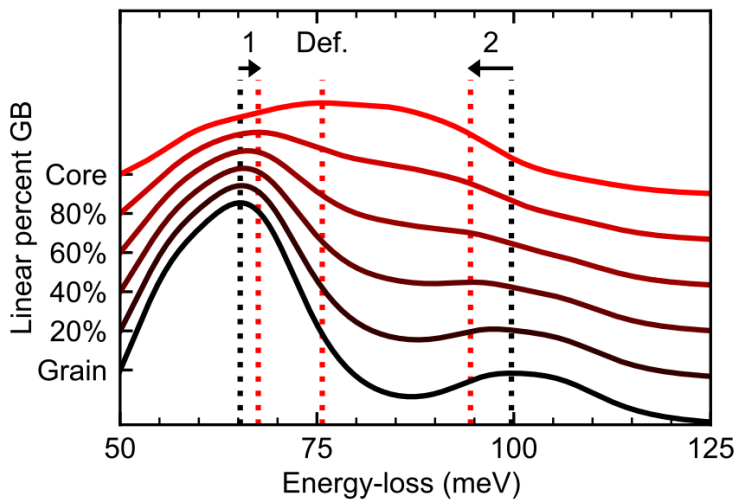

Figure S9. Profile of linear combination of grain- and intra-core PPhDOS with different ratio. From the bottom (grain) to the top (intracore), the ratio of intracore increases, resulting in the blue-shift of E1 and red-shift of E2, which is consistent with the EELS data in Figure 3(d).

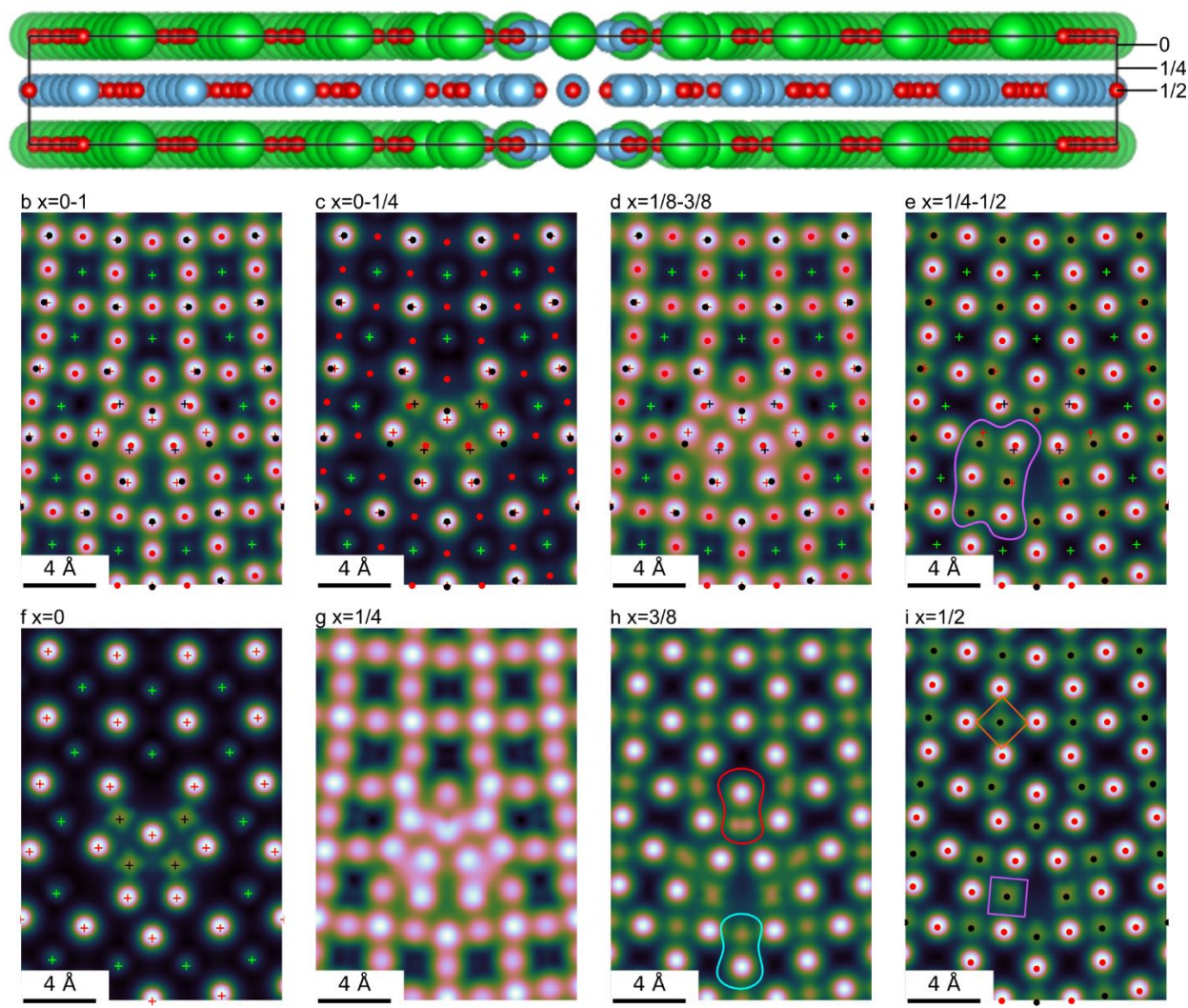

Figure S10. **Sections of DFT Charge Density at the Top Dislocation-core.** (a) Side view of the DFT supercell showing fractional coordinates of the supercell for integration. Integrated charge density from (b) 0-1 (c) 0-1/4, (d) 1/8-3/8, and (e) 1/4-1/2. Individual slices of charge density at (f) 0, (g) 1/4, (h) 3/8, and (i) 1/2. In (a-f,i) (green) Sr, (black) Ti, and (red) O atoms in the 0 plane are annotated with "+" markers while atoms in the 1/2 plane are annotated with "•" markers. Markers were excluded in (g,h) giving a cleaner unobscured image of the charge density between atomic planes. Purple annotations in (e,i) indicate rock-salt like TiO packing. Red annotations in (h) highlight anisotropically bonded Ti and O atoms. The cyan annotation in (h) highlights a region of increased orbital overlap. The orange annotation in (i) highlights the Ti$d$ orbitals in comparison to the rock-salt Ti$d$ orbitals annotated with purple.

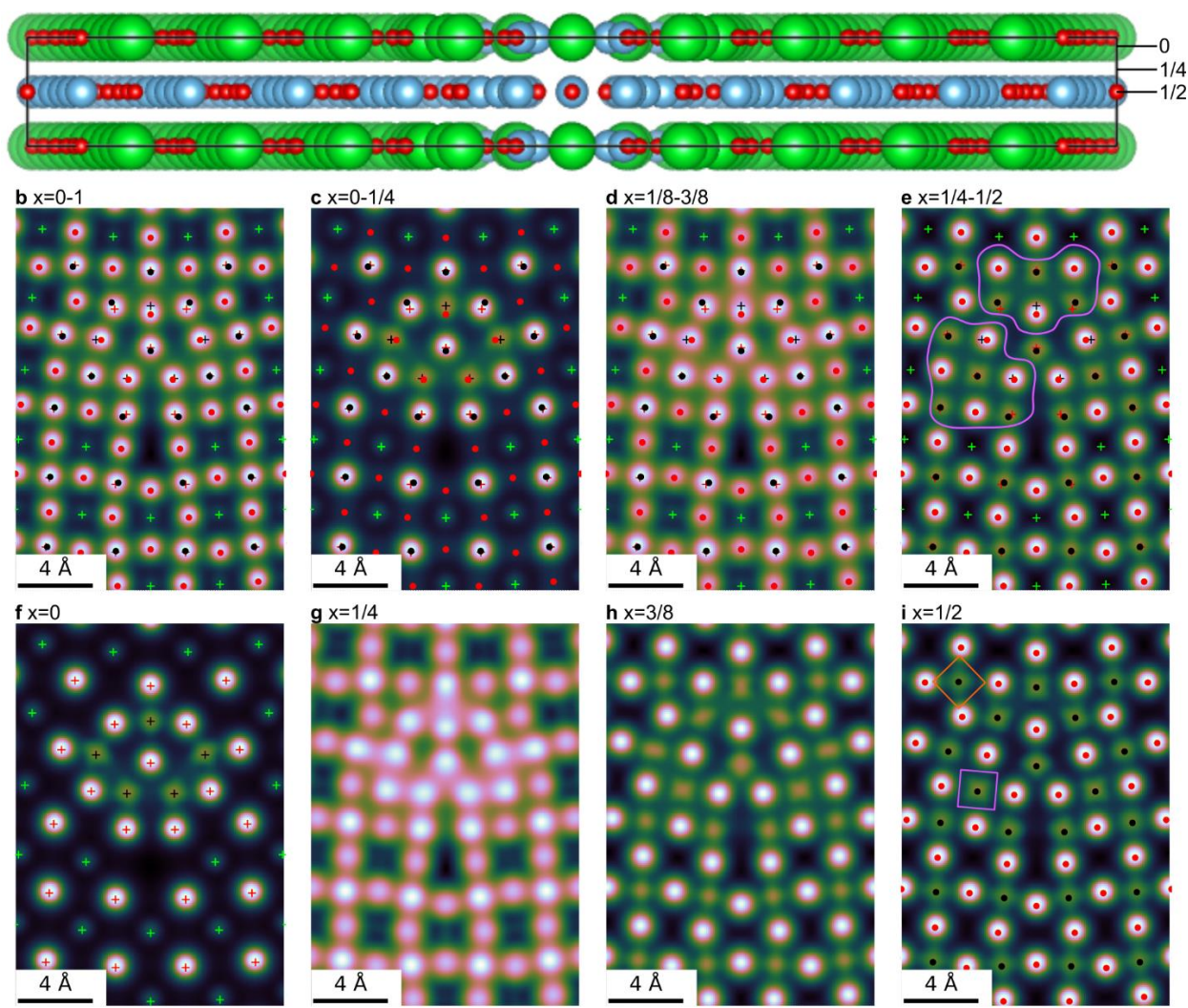

Figure S11. **Sections of DFT Charge Density at the Bottom Dislocation-core.** (a) Side view of the DFT supercell showing fractional coordinates of the supercell for integration. Integrated charge density from (b) 0-1 (c) 0-1/4, (d) 1/8-3/8, and (e) 1/4-1/2. Individual slices of charge density at (f) 0, (g) 1/4, (h) 3/8, and (i) 1/2. In (a-f,i) (green) Sr, (black) Ti, and (red) O atoms in the 0 plane are annotated with "+" markers while atoms in the 1/2 plane are annotated with "•" markers. Markers were excluded in (g,h) giving a cleaner unobscured image of the charge density between atomic planes. Purple annotations in (e,i) indicate rock-salt like TiO packing. The orange annotation in (i) highlights the Ti$d$ orbitals in comparison to the rock-salt Ti$d$ orbitals annotated with purple.

## S5 Supplemental Information: References

1. Egerton, R. F. *Electron Energy-Loss Spectroscopy in the Electron Microscope*. (Springer US, 2011).